\documentclass[preprint]{imsart}

\usepackage{enumerate}
\usepackage{amsthm,amsmath}
\usepackage{amssymb, amsfonts, stackrel}
\usepackage[authoryear,round,sort&compress]{natbib}
\usepackage{graphicx}
\usepackage{multirow}
\usepackage{float}
\usepackage{morefloats}
\usepackage{placeins}
\usepackage[list=true, listformat=simple]{subcaption}
\usepackage{color}

\newtheorem{thm}{Theorem}[section]
\newtheorem{corollary}{Corollary}[thm]
\newtheorem{lemma}{Lemma}[section]

\newcommand{\distributedas}{\overset{d}{=} }
\newcommand{\inftystar}{{\infty^*}}
\newcommand{\E}{ {\rm{E}} }

\usepackage{array}
\newcolumntype{L}[1]{>{\raggedright\let\newline\\\arraybackslash\hspace{0pt}}m{#1}}
\newcolumntype{C}[1]{>{\centering\let\newline\\\arraybackslash\hspace{0pt}}m{#1}}
\newcolumntype{R}[1]{>{\raggedleft\let\newline\\\arraybackslash\hspace{0pt}}m{#1}}

\startlocaldefs
\endlocaldefs

\begin{document}
\begin{frontmatter}
\title{Model-based curve registration via stochastic approximation EM algorithm}
\runtitle{Registration via SAEM}

\begin{aug}
\author{\fnms{Eric} \snm{Fu}\ead[label=e1]{ericfu@stat.ubc.ca}}
\and
\author{\fnms{Nancy} \snm{Heckman}\ead[label=e2]{nancy@stat.ubc.ca}}

\address{Department of Statistics, University of British Columbia, Vancouver, Canada\\
\printead{e1,e2}}
\end{aug}

\begin{abstract}
Functional data often exhibit both amplitude and phase variation around a common base shape, with phase variation represented by a so called warping function. The process of removing phase variation by curve alignment and inference of the warping functions is referred to as curve registration. When functional data are observed with substantial noise, model-based methods can be employed for simultaneous smoothing and curve registration. However, the nonlinearity of the model often renders the inference computationally challenging. In this paper, we propose an alternative method for model-based curve registration which is computationally more stable and efficient than existing approaches in the literature. We apply our method to the analysis of elephant seal dive profiles and show that more intuitive groupings can be obtained by clustering on phase variations via the predicted warping functions.
\end{abstract}

 \revised{\sday{18} \smonth{12} \syear{2017}}

\end{frontmatter}

\doublespacing

\section{Introduction} 
\label{sec:motivation}

  Functional data often exhibit  variation not only in amplitude but also in horizontal scaling, or phase.  
  For example, in the classical Berkeley Growth Study,  which considers growth as a function of age~\citep{james2006functional}, both the magnitude  of growth spurts and their age of occurrence vary across subjects.

  Early approaches to analysis of functional data  ignored phase variation, modelling data $(t_{ij}, y_{ij})$ for individual curves, $i=1, \ldots, N$, and sampled points, $j=1, \ldots, n_i$, as
  \begin{equation}
    \label{eq:fda}
    y_{ij} = f_i(t_{ij}) + \epsilon_{ij}
  \end{equation}
  with 
  $f_i$  a smooth function
  and $\epsilon_{ij}$  a noise term.
  The function $f_i$ is typically modelled flexibly, in terms of spline functions, either via minimizing a penalized likelihood \citep{brumback1998smoothing} or by a linear mixed effects model~\citep{rice2001nonparametric,durban2005simple,heckman2013penalized}.  Indeed, under general conditions, one can show that penalized likelihood estimates are the same as linear mixed effects model estimates~\citep{wand2003smoothing}.  \citet{yao2005functional} took a different approach by simply assuming that the $f_i$s were samples from a smooth stochastic process which can be approximated parsimoniously by the principal functions of the estimated covariance function.

  Often, however, data do not follow model (\ref{eq:fda}), but rather require inclusion of phase variation via a time transformation, $h_i$, also called a warping function:
  \begin{equation}{}
    y_{ij} = (f_i \circ h_i)(t_{ij}) + \epsilon_{ij}.
    \label{eq:functional-composition-model}
  \end{equation}
  The warping function  $h_i$ is continuous, strictly increasing and maps the sampling time, $t$, to the system time, $h_i(t)$, in such a way that, in system time, the $N$ curves  are in synchrony. 
  Ignoring phase variation and analyzing unaligned curves, that is, using model (\ref{eq:fda}) instead of model (\ref{eq:functional-composition-model}), can lead to incorrect conclusions.   For instance, estimating a population mean curve by the point-wise average of non-aligned individual curves may lead to dampened or even completely masked peaks and valleys.

  A common approach to analyzing data as in~(\ref{eq:functional-composition-model})  is to first smooth the $N$ curves individually and then align the curves to remove phase variations. 
  The aligned curves are then analyzed as if they are generated from~(\ref{eq:fda}).
  %
  %
  The process of curve smoothing and alignment is often called  curve registration.  
  Some well-known registration approaches are landmark registration \citep{kneip1992statistical,gasser1995searching,james2006functional} and methods based on optimization of some fitting criteria \citep{sakoe1978dynamic,ramsay1998curve,james2006functional,srivastava2011registration}. 
  Essentially, these procedures aim to align the peaks and valleys of the curves or their derivatives.

  This sequential approach  to fitting model (\ref{eq:functional-composition-model}) seems to work well when the observed data arise from smooth curves possessing a common set of clearly recognizable features.
  However, as pointed out by \citet{raket2014nonlinear}, the pre-smoothing step can be problematic for noisy data.
  In this case an individual curve's data might be overfitted, resulting in a ``bumpy" curve with fictitious features being used in the subsequent alignment.
  Another problem with the sequential approach is with inference on population level parameters, since the sequential approach does not take into account the uncertainty of the alignment.

  In this paper, we consider an alternative to the sequential approach and work directly with the model in (\ref{eq:functional-composition-model}), estimating the amplitude and phase variation simultaneously via a new expectation-maximization (EM) algorithm.  
  Our proposed model is similar to those of \citet{brumback2004self}, \citet{telesca2008bayesian} and \citet{raket2014nonlinear}. 
  In Section 2, we discuss and constrast our model to the existing ones in the literature.
  We also discuss the identifiablity of the models, providing an easy to apply theorem, similar to that in \citet{chakraborty2017functional}.
  Through a simulation study, we compare all four approaches, finding our EM based method and the Bayesian hierarchical curve registration (BHCR) of \citet{telesca2008bayesian} computationally more stable, especially when the curves are densely observed. 
  In terms of statistical efficiency, the estimated base shape and predicted warping functions using our proposed method also have the lowest mean squared errors in general.
  We then  apply our method to the analysis of dive depth trajectories of elephant seals recorded by biologging devices at high sampling frequency. 
  For each dive, we estimate both the amplitude parts, $f_i$s and warping functions, $h_i$s.  
  The estimated warping functions are then used to cluster the depth trajectories into groups of similar shape.
  We find that clustering based  on the warping function is superior to clustering based on the original data, yielding more interpretable clusters.

\section{Model-based registration} 
\label{sec:model_based_registration}

  In this section, we discuss our proposed model for~(\ref{eq:functional-composition-model}) and contrast with similar models suggested by \citet{brumback2004self}, \citet{telesca2008bayesian} and \citet{raket2014nonlinear}. Following that, we also discuss the slightly different approach of \citet{raket2014nonlinear}.

  The first three methods extend the shape-invariant model (SIM) of~\citet{lawton1972self}, where $f_i$ is a vertical shifting and scaling of a common base shape $f$:
  \begin{equation}
    f_i(t) = a_{i,sh} + a_{i,sc} f(t)
    \label{eq:rank-1-amplitude-function}  
  \end{equation}
  where $a_{i,sh}$ and $a_{i,sc}$ are Gaussian random effects with means $\mu_{sh}=0$ and $\mu_{sc}=1$, and $f$ is modelled as a linear combination of B-spline basis functions, as
  \begin{equation}
    f(t) = \sum_{k=1}^{K^f} \alpha_k B^{f}_k(t).
    \label{eq:B-spline_common_shape}
  \end{equation}
  In our method and that of \citeauthor{brumback2004self}, the $\alpha_k$s are fixed.
  \citeauthor{telesca2008bayesian} work in a hierarchical Bayesian framework where $\alpha_k$s are random.   

  In the first three methods, the warping function $h_i$ is modelled as random and strictly monotone and continuous, using B-spline basis functions:
  \begin{equation}
    h_i(t) = \sum_{k=1}^{K^h} \beta_{i,k}B^h_k(t)
    \label{eq:random-time-transformation}
  \end{equation}
  with the $\beta_{i,k}$s random and $\E[h_i(t)] = t$ for all $t$.
  To ensure that $h_i$ is a proper warping function that maps onto $[0, 1]$, the models use the fact that $h_i$ is strictly increasing if $\beta_{i,1} < \beta_{i,2} < \cdots < \beta_{i,K_h}$ \citep{kelly1990monotone} and force $h_i(0)=0$ and $h_i(1)=1$ by setting $\beta_{i,1}=0$ and $\beta_{i,K^h}=1$. 
  While \citeauthor{telesca2008bayesian} explicitly impose the constraint on $\beta_{i,k}$s which is otherwise multivariate Gaussian, \citeauthor{brumback2004self} link the basis coefficients to Gaussian warping effects by the Jupp transformation \citep{jupp1978approximation}.

  We take a more straightforward but non-Gaussian approach. We model the first difference of the $\beta_{i,k}$s by a Dirichlet distribution.
  Specifically, we express the coefficients of the warping functions as
  \begin{equation}
    \beta_{i,k} = \sum_{j=1}^{k}w_{i,j} 
    \label{eq:cumsum-warping-coefficient}  
  \end{equation}
  with
  \begin{equation}
    w_{i,1}\equiv 0, ~~~~~ \tilde{\bf w}_{i} \equiv (w_{i,2}, \ldots, w_{i,K^h})^\top \sim \text{Dirichlet}({\boldsymbol\kappa}_0, \tau).
    \label{eq:random-warping-distribution}
  \end{equation}
  The Dirichlet distribution ensures that $w_{i,1} < w_{i,1} + w_{i,2} < \cdots < w_{i,1} + \cdots + w_{i,K^{h}} = 1$.
  Since $h_i(t)$ is linear in $w_{i,j}$s, our model allows easy and exact control on $\E[h_i(t)]$.
  Also, there is only one unknown parameter, $\tau$, to be estimated, regardless of the number of coefficients in the spline in~(\ref{eq:random-time-transformation}).
  Although the correlation structure of this Dirichlet distribution is fixed, we have seen in simulation studies that the conditional expectation of the warping function given the data provides a good estimate of the true warping function when the number of observations per curve is large.

  \citet{raket2014nonlinear} proposed a different type of model for warped curves. 
  The data are modelled as $y_{ij} = f(h_i(t_{ij})) + x_i(t_{ij}) + \epsilon_{ij}$.
  The common base shape, $f$, is non-random and subject to warping, whereas $x_i$ a zero mean Gaussian process representing idiosyncratic features that will not be aligned.
  The base shape, $f$, is represented by an interpolating spline over all distinct $t_{ij}$s while the warping function, $h_i$, is a realization of a Brownian bridge.
  Although there is no constraint discussed for monotonicity of $h_i$ in their original paper, their actual implementation allows setting $h_i$ to an interpolating spline through the realized Brownian bridge with monotonicity ensured by Hyman filtering.

  \subsection{Identifiability}  

  In general, model (\ref{eq:functional-composition-model}) is not identifiable. 
  However, we can show that if $f_i$ is a random shifting and scaling of a base shape, as in our proposed model, identifiability can be restored under some additional conditions.
  We use the notation $X_1 \distributedas X_2$ to mean that $X_1$ and $X_2$ have the same  distribution.
  Let $\tilde{f}_i = f_i \circ h_i$ for $i=1, 2$.

  \begin{thm}
    \label{thm:identifiable}
    Suppose that 
    \begin{itemize}
    \item[1)] 
      $$f_i(t) = a_{i,sh} + a_{i,sc} \xi_i(t)$$ where $a_{i,sh}$ and $a_{i,sc}$ are random with $\text{P}(a_{i,sc}=0)=0$ and $\xi_i$, defined on $[0,1]$, is a real-valued non-random function with a continuous first derivative which vanishes at most on a countable subset of $[0,1]$; and
    \item[2)] with probability one,  $h_i$ is continuous with strictly positive and continuous first derivative, with $h_i(0)=0$ and $h_i(1)=1$; and
    \item[3)] $\E(h_i^{-1}(t))=t ~~~~ \forall t$.
    \end{itemize}
    Then
    \[
    \tilde{f}_1 \distributedas \tilde{f}_2 ~~ {\rm{if~and~only~if}} ~~  (f_1, h_1)  \distributedas (f_2, h_2).
    \]
  \end{thm}

  \begin{corollary}
    \label{cor:identifiable.1}
    Assume that the conditions of Theorem \ref{thm:identifiable} 
    hold.  Suppose also that, for $i=1,2$,  $(a_{i,sh},a_{i,sc})$ has known mean $(\mu_{sh},\mu_{sc})$ with $\mu_{sc} \neq 0$.  If $\tilde{f}_1 \distributedas \tilde{f}_2$, then
    $h_1 \distributedas h_2$, $\xi_1 = \xi_2$ and $(a_{1,sh},a_{1,sc}) \distributedas (a_{2,sh},a_{2,sc})$.
  \end{corollary}

  \begin{corollary}
    \label{cor:identifiable.2}
    Assume that the conditions of Theorem \ref{thm:identifiable}  hold.  Suppose also that, for $i=1,2$,  $a_{i,sh} \equiv 0$ and  $\int_t \xi_i^2(t)~{\rm{d}}t=1$.  
    If $\tilde{f}_1 \distributedas \tilde{f}_2$, then
    $h_1 \distributedas h_2$, $\xi_1 (t)= \pm \xi_2(t)$ and $a_{1,sc} \distributedas \pm a_{2,sc} $.
  \end{corollary}

  The proof of the theorem essentially follows the proof of Theorem 1 of \citet{chakraborty2017functional} who introduce the use of local variation.  These authors restrict themselves to  the case that $a_{i,sh}=0$ and directly obtain the result  given in Corollary~\ref{cor:identifiable.2}. 
  Given their result, our theorem is not surprising since adding the constant $a_{i,sh}$ does not change the local variation functions used in the proof.
  We sketch our proof  of the theorem in the Appendix, including more details of the topology of the function space considered and, in some cases, with  modified and shortened   arguments.  The proofs of the corollaries are straightforward, with the proof of Corollary~\ref{cor:identifiable.2} following the ending arguments of \citet{chakraborty2017functional}.

  \vskip 10pt
  Note that the theorem's restriction on $h$ is E($h_i^{-1}(t))=t$ and that the method of proof of  Chakraborty and Panaretos relies heavily on this assumption.  
  However, for computational simplicity, we assume instead that E$(h_i(t))=t$.   With this alternate assumption, our method still provides good estimates of the base and warping functions, as indicated in our simulation studies.  We note that \citet{brumback2004self} and  \citet{telesca2008bayesian}   make the same assumption as we do.

  All three methods fix the mean of $a_{i,sh}$ and $a_{i,sc}$ as in Corollary~\ref{cor:identifiable.1} with $\mu_{sh}=0$ and $\mu_{sc}=1$.
  In addition, \citeauthor{brumback2004self} restrict the $a_{i,sh}$s and ${a}_{i,sc}$s by using  rank deficient variance-covariance structures to guarantee that the sample means of the random effects are equal to their known population means. 
  Although the concept of identifiability is not as well defined in the Bayesian context, to improve the computational stability of the posterior inference, \citet{telesca2008bayesian} do adjust the sampled amplitude random effects in each MCMC steps such that the sample means of the shifting and scaling factor are equal to zero and one respectively.

  To estimate the model parameters, \citet{brumback2004self}  and \citet{raket2014nonlinear} modify the algorithm of \citet{lindstrom1990nonlinear} for maximum likelihood estimation.
  \citet{telesca2008bayesian} take a Bayesian approach with MCMC-based posterior inference.
  In the following section, we proposed an EM algorithm with stochastic approximation for parameter estimation.
  The algorithm also conveniently yield the fitted curves and the prediction warping functions as a by-product of the stochastic approximation.

\section{Estimation} 
\label{sec:estimation}

  Recall that our model~(\ref{eq:rank-1-amplitude-function}) has a non-random base function $f$, parameterized by the basis coefficients ${\boldsymbol\alpha} = (\alpha_1,\ldots, \alpha_{K^f})$ as in (\ref{eq:B-spline_common_shape}),  random shifting and scaling effects ${\bf{a}}_i \equiv (a_{i,sc}, a_{i,sh})$ which are bivariate normal with fixed mean ${\boldsymbol\mu}_0 = (\mu_{sh}, \mu_{sc})^\top = (0, 1)^\top$ and unknown covariance $\Sigma$, and random warping functions based on ${\bf{w}}_i$s, which are Dirichlet with parameter $\tau$ as in (\ref{eq:random-warping-distribution}).  We use the following EM algorithm for maximum likelihood estimation of ${\boldsymbol\alpha}$, $\Sigma$, $\tau$ and error variance $\sigma^2$ of the proposed model.
  For the $i^\text{th}$ curve, $i=1, \ldots, N$, denote the observed data by ${\bf y}_i$
  and  the complete data by ${\bf z}_i = ({\bf y}_i, {\bf a}_i, {\bf w}_i)$.  Let  ${\boldsymbol\theta} = ({\boldsymbol\alpha}, {\sigma^2}, {\Sigma}, {\tau})$ be the collection of all unknown parameters and $ \ell_c(\theta;{\bf z})$  the complete-data log-likelihood function.
  To find the maximum  likelihood estimate of  ${\boldsymbol\theta}$, the EM algorithm \citep{dempster1977maximum} maximizes the observed data likelihood by creating a sequence, $\{\boldsymbol\theta^{(k)}, k \geq 1 \}$, via iterations between 1) an E-step that computes $Q({\boldsymbol\theta}; {\boldsymbol\theta}^{(k)}) = \mathbb{E}\left[\ell_c(\boldsymbol\theta;{\bf z})\middle|{\bf y};\boldsymbol\theta^{(k)}\right]$ using $\boldsymbol\theta^{(k)}$ as the true parameter value, and 2) an M-step that sets $\boldsymbol\theta^{(k+1)} = \arg\max Q(\boldsymbol\theta;\boldsymbol\theta^{(k)})$.

  The complete-data log-likelihood consists of three components,
  \begin{equation*}
    \ell_c(\theta;{\bf z}) = 
    \ell^{\bf a}_c(\Sigma;{\bf a}) +
    \ell^{\bf w}_c(\tau;{\bf w}) + 
    \ell^{\bf y}_c({\boldsymbol\alpha}, \sigma^2;{\bf z})
  \end{equation*}
  where  the log-likelihoods of the random effects ${\bf a}=$$({\bf a}_1, \ldots {\bf a}_N)$ and ${\bf w} = $$({\bf w}_1,\ldots,{\bf w}_N)$ are
  \begin{equation}
    \begin{aligned}
      \ell^{\bf a}_c(\Sigma;{\bf a}) 
      & = -N\log(2\pi) - \frac{N}{2}\log\det\Sigma - \frac{1}{2}\sum_{i=1}^{N}({\bf a}_i - {\boldsymbol\mu}_0)^\top\Sigma^{-1}({\bf a}_i - {\boldsymbol\mu}_0) \\
      & = -N\log(2\pi) + -\frac{N}{2}\log\det\Sigma -\frac{1}{2} \text{tr}\left(\sum_{i=1}^{N}({\bf a}_i-{\boldsymbol\mu}_0)({\bf a}_i-{\boldsymbol\mu}_0)^\top\Sigma^{-1}\right)
    \end{aligned}
    \label{eq:complete-data-likelihood-a}
  \end{equation}
  and
  \begin{equation}
    \ell^{\bf w}_c(\tau;{\bf w}) = \sum_{k=2}^{K_h}(\tau\kappa_k-1)\sum_{i=1}^{N}\log(w_{i,k} - w_{i,k-1}) - N\left(\sum_{k=2}^{K_h}\log\Gamma(\tau\kappa_k) - \log\Gamma(\tau)\right)
    \label{eq:complete-data-likelihood-w}
  \end{equation}
  and the conditional normal log-likelihood of the observed curves given the random effects is
  \begin{equation}
    \begin{aligned}
      \ell^{\bf y}_c({\boldsymbol\alpha}, \sigma^2;{\bf z}) 
      & = -\frac{n_{\text{tot}}}{2} - \frac{1}{2\sigma^2}\sum_{i=1}^{N} ||{\bf y}_i - a_{i,sh}{\bf 1} - a_{i,sc}B_i({\bf w}_i){\boldsymbol\alpha}||^2\\
      & = -\frac{n_{\text{tot}}}{2}\log(2\pi\sigma^2) - \frac{1}{2\sigma^2}
      \left\{
      \sum_{i=1}^{N}||{\bf y}_i - a_{i,sh}{\bf 1}||^2 \right.\\
      & ~~~~~ - 2\sum_{i=1}^{N}({\bf y}_i - a_{i,sh}{\bf 1})^\top (a_{i,sc}B_i({\bf w}_i))\boldsymbol\alpha \\
      & ~~~~~ + \left. \text{tr}\left(\sum_{i=1}^{N}a_{i,sc}^2B_i({\bf w}_i)^\top B_i({\bf w}_i){\boldsymbol\alpha\boldsymbol\alpha^\top}\right)
      \right\},
    \end{aligned}
    \label{eq:complete-data-likelihood-y}
  \end{equation}
  with $n_{\text{tot}}$ the total number of observed values and 
  \begin{equation}
    B_i({\bf w}_i) = 
    \left(
    \begin{array}{ccc}
      B^f_1(h(t_{i,1}; {\bf w}_i)) & \cdots & B^f_K(h(t_{i,1}; {\bf w}_i))\\
      \vdots & \ddots & \vdots \\
      B^f_1(h(t_{i,n_i}; {\bf w}_i)) & \cdots & B^f_K(h(t_{i,n_i}; {\bf w}_i))\\
    \end{array}
    \right)
    \label{eq:B-spline-basis-eval-matrix}
  \end{equation} 
  a basis evaluation matrix for the shape function at warped times $h(t_{i,1};{\bf w}_i), \ldots, h(t_{i,n_i};{\bf w}_i)$ for curve $i$.

  For our model, we can see from~(\ref{eq:complete-data-likelihood-a})--(\ref{eq:complete-data-likelihood-y}) that the complete-data log-likelihood is linear in sufficient statistics:
  \begin{equation*}
    \begin{array}{lclcl}
      S_{{\bf y}{\bf y}}&=& \displaystyle\sum_{i=1}^{N}S_{{\bf y}{\bf y},i}&\equiv&\displaystyle\sum_{i=1}^{N}({\bf y}_i-a_{i,sh}{\bf 1})^\top({\bf y}_i-a_{i,sh}{\bf 1}),\\[20pt]
      S_{B{\bf y}}   &=& \displaystyle\sum_{i=1}^{N}S_{B{\bf y},i}   &\equiv&\displaystyle\sum_{i=1}^{N}a_{i,sc} B_i({\bf w}_i)^\top({\bf y}_i-a_{i,sh}{\bf 1}),\\[20pt]
      S_{BB}      &=& \displaystyle\sum_{i=1}^{N}S_{{BB},i}    &\equiv&\displaystyle\sum_{i=1}^{N}a_{i,sc} B_i({\bf w}_i)^\top B_i({\bf w}_i)a_{i,sc},\\[20pt]
      S_{\bf a}         &=& \displaystyle\sum_{i=1}^{N}S_{{\bf a},i }      &\equiv&\displaystyle\sum_{i=1}^{N}({\bf a}_i-{\boldsymbol\mu}_0)({\bf a}_i-{\boldsymbol\mu}_0)^\top\text{ and}\\[20pt]
      S_{{\bf w}_k}     &=& \displaystyle\sum_{i=1}^{N}S_{{\bf w}_k,i}     &\equiv&\displaystyle\sum_{i=1}^{N}\log(w_{i,k+1}-w_{i,k}) \text{ for } k=2\ldots, K_h.
    \end{array}
    \label{eq:sufficient-statistics}
  \end{equation*}
  Therefore  $ Q(\boldsymbol\theta;\boldsymbol\theta^{(k)})$ depends on the observed data only through the conditional expectation of these sufficient statistics.
  Let $S$ be a generic notation for the sufficient statistics and $\widetilde{S}^{(k)}$ its conditional expectation given the observed data under the ``true" parameter $\theta^{(k)}$. 

  Given the $\tilde{S}^{(k)}$'s, the M-step is relatively straightforward. Closed-form solutions exist for updating the estimates of $\boldsymbol\alpha$, $\Sigma$ and $\sigma^2$:
  \begin{equation*}
    \begin{aligned}
      \boldsymbol\alpha^{(k+1)} & = \widetilde{S}_{BB}^{-1}\widetilde{S}_{B {\bf y}}\\
      \boldsymbol\Sigma^{(k+1)} & = \frac{1}{N}\widetilde{S}_{{\bf a}}\\
                       {\boldsymbol\sigma^2}^{(k+1)} & = \frac{1}{n_\text{tot}}\left(\widetilde{S}_{{\bf yy}} - 2\widetilde{S}_{B {\bf y}}^\top\boldsymbol\alpha^{(k+1)} + {\boldsymbol\alpha^{(k+1)}}^\top\widetilde{S}_{BB}\boldsymbol\alpha^{(k+1)}\right).
    \end{aligned}
    \label{eq:M-step}
  \end{equation*}
  For the concentration parameter of the Dirichlet warping effects, the corresponding maximizer,
  \begin{equation*}
    \tau^{(k+1)} = \underset{\tau>0}{\arg\max} \left\{ \sum_{k=2}^{K_h}(\tau\kappa_{k} - 1)\widetilde{S}_{{\bf w}_k} - 
    N\left(\sum_{k=2}^{K_h}\log\Gamma(\tau\kappa_{k}) - \log\Gamma(\tau)\right) \right\},
    \label{eq:M-step-tau}
  \end{equation*}
  can be solved numerically by Newton's methods.

  \subsection{The E-step}

  Calculating the conditional expectations, the $\tilde{S}^{(k)}$s, is difficult, making the E-step challenging.  Explicit calculation of these  conditional expectations practically impossible because of the model's non-linearity, caused by the warping functions.   Thus sampling-based methods are often employed to approximate the expectations.  \citet{wei1990monte} considered Monte Carlo approximations with direct sampling from the distribution of $({\bf a}_i, {\bf w}_i) | {\bf y}_i;{\boldsymbol\theta}^{(k)}$.  However, in our case, the conditional distribution  is intractable.  
  We tried the importance sampling approach of \cite{walker1996algorithm} but we found the efficiency of the importance sampler low for generating samples of ${\bf w}_i$s; 
  the importance weights were concentrated only on a few points, resulting in a small effective sample size.
  This is not surprising given the high sampling freuency; the conditional density of the warping functions is likely to be concentrated in a small set.

  We also consider approximating $\tilde{S}^{(k)}$ by Markov Chain Monte Carlo (MCMC) where at the $k\text{th}$ step of the EM algorithm, for each curve, we sample a sequence of random effects, $\left\{({\bf a}^{(k)}_{i,[r]}, {\bf w}^{(k)}_{i,[r]}); r=1, \ldots, R_k \right\}$, by a Metropolis-Hastings algorithm with $({\bf a}_i, {\bf w}_i) | {\bf y}_i;{\boldsymbol\theta}^{(k)}$ as the stationary distribution.  
  We then approximate conditional expectations  by ergodic averages,
  \begin{equation*}
    \hat{S}_{MC, i}^{(k)} = \frac{1}{R_k}\sum_{r=1}^{R_k} S\left({\bf a}^{(k)}_{i,[r]}, {\bf w}^{(k)}_{i,[r]}, {\bf y}^{(k)}_{i,[r]}\right).
  \end{equation*}
  However, this method is computationally too intensive, largely due to the evaluation of a B-spline basis matrix in (\ref{eq:B-spline-basis-eval-matrix}) for each MCMC sample of warping effect, ${\bf w}^{(k)}_{i,[r]}$.  In addition,  
  our experience shows that $R_k$ must be large for accurate approximation of the expectations.
  Indeed, in theory, the MCMC sample size, $R_k$, must increase with $k$ so that the approximation error does not dominate and the EM algorithm can converge~\citep{wei1990monte}.
  %

  \cite{kuhn2005maximum} modify this method to avoid generating a large MCMC sample afresh at each EM iteration, and we apply their modification here.
  We set 
  \begin{equation}
    \hat{S}^{(k)}_{SA, i} = \hat{S}_{SA, i}^{(k-1)} + \gamma_k\left(\hat{S}_{MC, i}^{(k)} - \hat{S}_{SA, i}^{(k-1)}\right)
    \label{eq:stochastic-approximation-scheme}
  \end{equation}
  which, instead of discarding all samples from previous E-steps, updates $\hat{S}_{SA, i}^{(k-1)}$, the conditional expectation approximation from the preceeding E-step, using $\hat{S}_{MC, i}^{(k)}$ calculated from the $k^{\text{th}}$ MCMC sample.
  \citeauthor{kuhn2004coupling} showed that if the step size, $\gamma_k$, satisfies 1) $0 \leq \gamma_k \leq 1$, 2) $\sum \gamma_k = \infty$ and 3) $\sum \gamma_k^2 < \infty$, and the Markov chains are uniform ergodic, then their stochastic approximation EM (SAEM) algorithm converges almost surely to a local maximum of the observed-data log-likelihood under some general conditions.
  In addition, the convergence of SAEM does not depend on $R_k$, which implies that a precise MCMC approximation of $\hat{S}_{MC, i}^{(k)}$ for each $k$ is not necessary.

  We use the SAEM, choosing the step size to be 
  \begin{equation*}
    \gamma_k = \left\{\begin{array}{cc}
    1 & \text{for } k \leq B\\
    (k-B)^{-\alpha} & \text{for }k > B
    \end{array}\right.
  \end{equation*}
  for $0.5 < \alpha \leq 1$ so that the first $B$ steps are burn-in steps where the SAEM algorithm can move to the vicinity of the maximum.
  For the MCMC updates, we choose $R_k\equiv1$ to minimize the number of B-spline basis evaluations at each step.

  We generate the Markov chains by a Metropolis-Hastings-within-Gibbs algorithm to sample ${\bf a}_i$ and ${\bf w}_i$ in turns.
  Sampling from the conditional distribution of  ${\bf a}_i$ is straightforward, as follows.  Since distribution of the observations given the random effects is
  \begin{equation*}
    {\bf y}_i | {\bf a}_i, {\bf w}_i \sim N(F_i({\bf w}_i){\bf a}_i, \sigma^2I)
  \end{equation*}
  where
  \begin{equation*}
    F_i({\bf w}_i) = \left(\begin{array}{cc}
      1 & f(h(t_{i,1}; {\bf w}_i))\\
      \vdots & \vdots\\
      1 & f(h(t_{i,n_i};{\bf w}_i))\\
    \end{array}
    \right)
  \end{equation*}
  and ${\bf a}_i$ is normally distributed with mean ${\boldsymbol \mu}_0$ and covariance $\Sigma$, the conditional distribution of ${\bf a}_i$ given ${\bf y}_i$ and ${\bf w}_i$ is also normal with covariance matrix $\Sigma_i = \left(\sigma^{-2}F_i({\bf w}_i)^\top F_i({\bf w}_i) + \Sigma^{-1}\right)^{-1}$ and mean $\boldsymbol\mu_i = \Sigma_i\left(\sigma^{-2}F_i({\bf w}_i)^\top{\bf y}_i + \Sigma^{-1}\boldsymbol\mu_0\right)$. 
  Therefore, The MCMC sample of ${\bf a}_i$ can be generated directly by a Gibbs sampler.

  On the other hand, the conditional distribution of ${\bf w}_i | {\bf a}_i, {\bf y}_i$ does not belong to a common family and direct sampling is difficult.
  Therefore we replace the Gibbs step by the following Metropolis-Hastings step. 
  First, we transform $\tilde{\bf w}_{i}$ from the $K^h-2$ simplex, $\mathbb{S}$, to the space $\mathbb{P} = \{{\bf x}\in\mathbb{R}^{K_h-1}: {\bf v}^\top{\bf 1} = 0\}$ by the centered-log-ratio transform, $[\mathcal{G}(\tilde{\bf w}_i)]_k = \log(w_{k+1}) - \left[\sum_{j=2}^{K_h}\log(w_j)\right]\big/\left(K^h-1\right)$.
  A proposal is drawn by performing a random walk on $\mathbb{P}$ where the random step follows a $(K^h-1)$-dimensional normal distribution with zero mean and a rank $K^h-2$ covariance matrix with diagonal elements equal to $\sigma^2_q\cdot (K^h-2)/(K^h-1)$ and off-diagonal elements equal to $-\sigma^2_q/(K^h-1)$ such that the sum-to-zero constraint is satisfied.
  The proposal is mapped from $\mathbb{P}$ back to $\mathbb{S}$ by the softmax transform, $\left[\mathcal{G}^{-1}({\bf x})\right]_{j+1} = e^{x_j} \big/ \sum_{k=1}^{K^h-1} e^{x_k}$. 
  The proposal for the warping effect, denoted by ${\bf w}^*$, is then accepted with probability,
  \begin{equation*}
    \min\left\{
    1, 
    \frac{f_{{\bf y}|{\bf a}, {\bf w}} \left(a^{(k)}_{i}, {\bf w}^* \middle| {\bf y}_i; \boldsymbol\theta^{(k)}\right)}
         {f_{{\bf y}|{\bf a}, {\bf w}} \left(a^{(k)}_{i}, {\bf w}^{(k-1)}_{i} \middle| {\bf y}_i; \boldsymbol\theta^{(k)}\right)}
         \cdot 
         \prod_{l=2}^{K_h}\frac{\left[{\bf w}^*\right]_l}{\left[{\bf w}^{(k-1)}_{i}\right]_l}
         \right\};
  \end{equation*}
  otherwise, the sampled state from the previous iteration is taken as the new state.
  In practice, we run each chain in the E-step with 5 to 10 iterations to encourage a better approximation to the stationary distribution. 
  This is in keeping with~\cite{kuhn2005maximum}, who found that a small number of burn-in iterations usually suffices and a longer burn-in does not improve the convergence of the SAEM algorithm much.

  To predict the warping function and the warped curves from the fitted model, we suggest using the conditional mean of the desired functions:
  \begin{equation*}
    \hat{y}_i(t)
    = 
    \text{E}\left(a_{i,sh}\middle|{\bf y}_i; \hat{\boldsymbol\theta}\right) 
    +
    \sum_{k=1}^{K^{f}}\text{E}\left(a_{i,sc}B^f_k(h(t;{\bf w}_i))\middle|{\bf y}_i; \hat{\boldsymbol\theta}\right) \hat\alpha_k
  \end{equation*} 
  and 
  \begin{equation*}
    \hat{h}_i(t) 
    = 
    \sum_{k=1}^{K^h}\sum_{j=1}^{k}\text{E}(w_{i,j}|{\bf y}_i; \hat{\boldsymbol\theta})B_k^h(t).
  \end{equation*}
  Stochastic approximations to the required conditional expectations are be easily obtained as a by-product of the SAEM algorithm.
  %

\section{Simulations} 
\label{sec:simulations}

  We compare the statistical efficiency of our proposed model for estimating the base shape and predicting the warping functions and the computational feasibility of the SAEM algorithm to existing model-based registration methods in the literature by a simulation study. 
  Data are simulated from the model discussed in Section 2 where the cubic splines for the base shape and the warping functions have equally spaced knots over the interval of $[0, 1]$.

\begin{figure}[ht]
    \centering
    \includegraphics[width=0.9\textwidth]{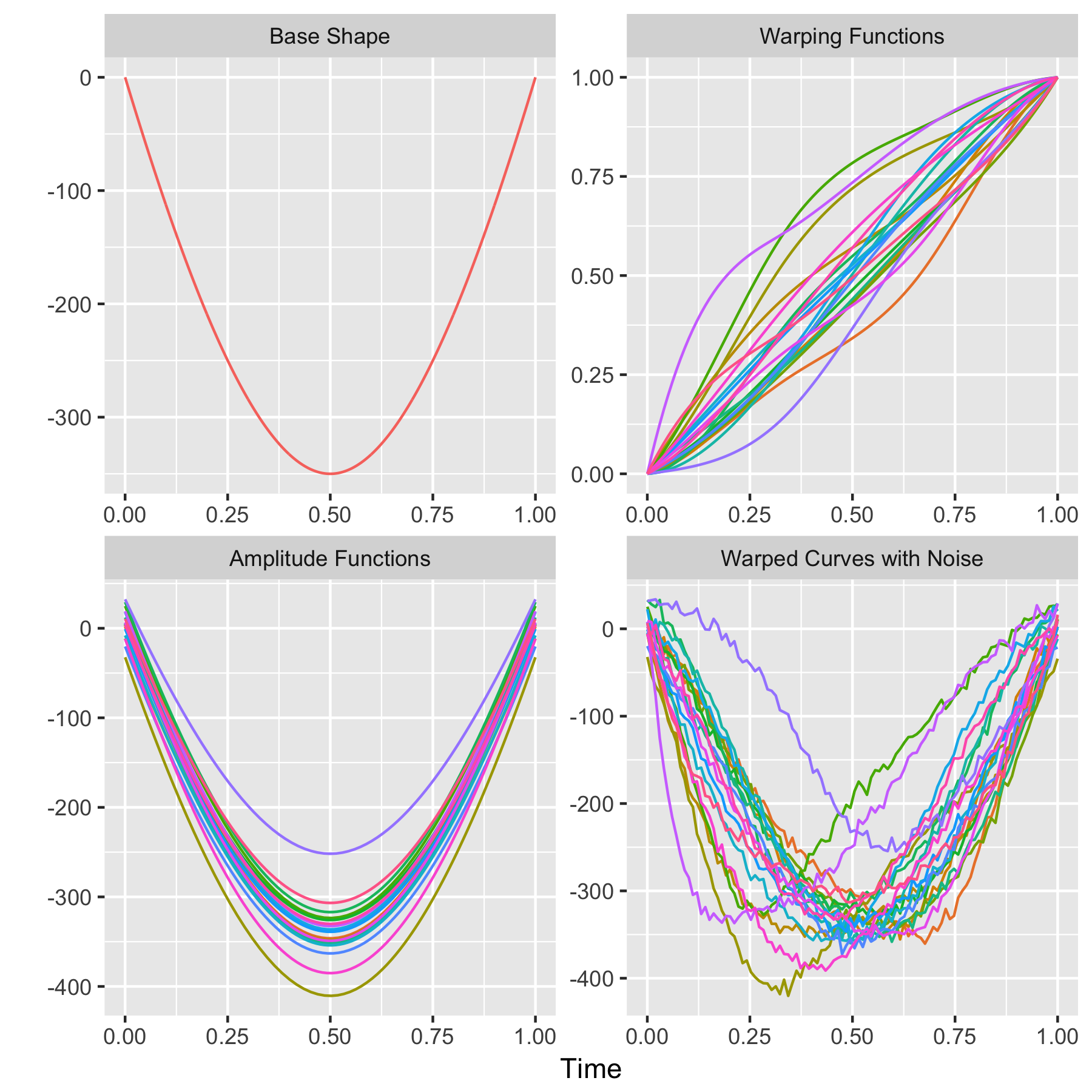}
    \caption{Twenty simulated curves of shape 1. Number of observations per curve is $n = 100$.}
    \label{fig:shape-1}
  \end{figure}

\begin{figure}[ht]
    \centering
    \includegraphics[width=0.9\textwidth]{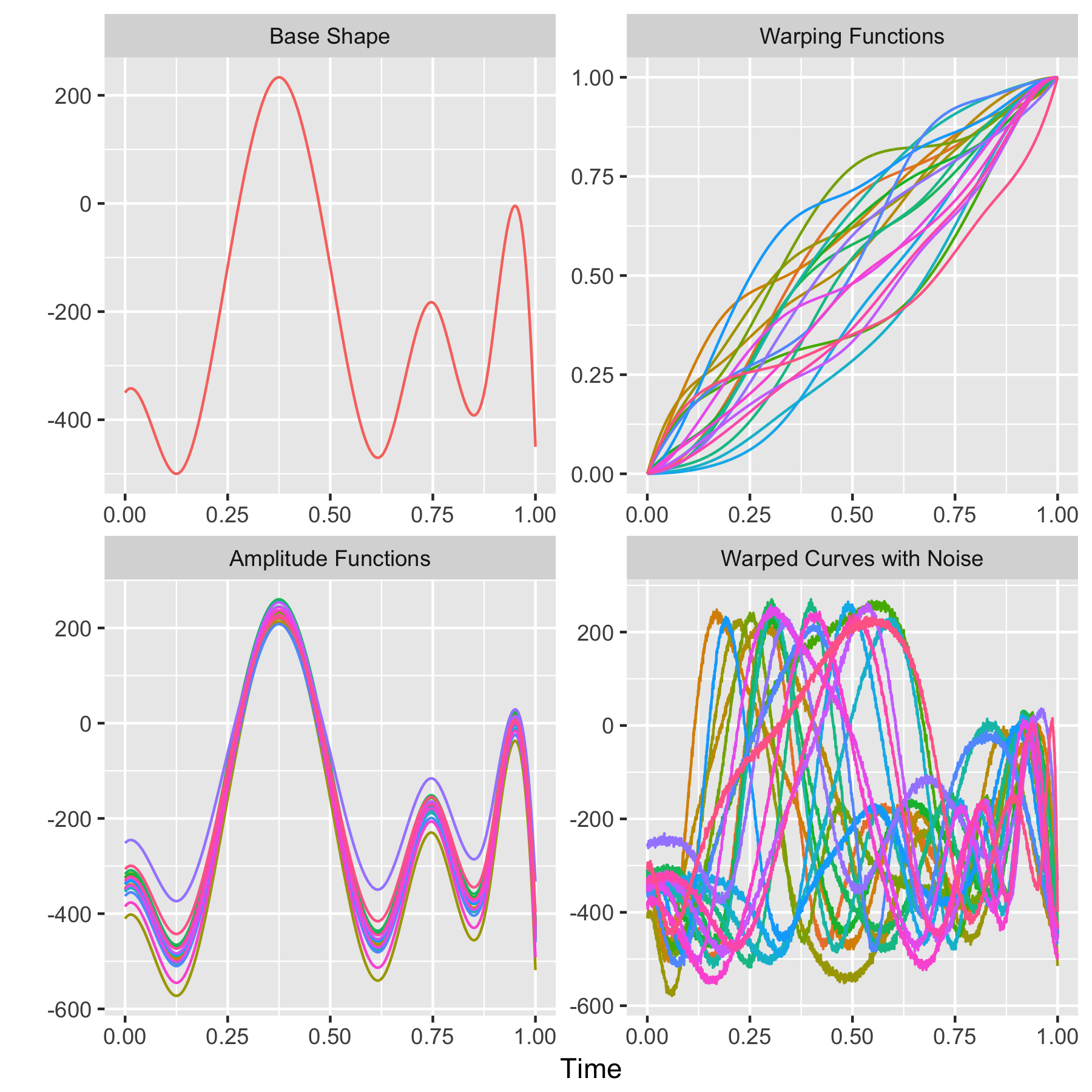}
    \caption{Twenty simulated curves of shape 2. Number of observations per curve is $n = 1000$.}
    \label{fig:shape-2}
  \end{figure}

  Two scenarios are considered.
  In the first case, the base shape uses 5 B-spline basis functions with coefficients equal to $0$, $-200$, $-500$, $-200$ and $0$ while the warping functions use 6 B-spline basis functions.
  In the second case, the base shape uses 11 B-spline basis functions with coefficients equal to $-350$, $-300$, $-700$, $-100$,  $400$, $-100$, $-700$,  $100$, $-800$,  $400$ and $-450$, while the warping functions use 9 B-spline basis functions.
  For the random effect distributions, in both cases, the concentration parameter of the Dirichlet warping effects is $\tau = 10$ and the covariance matrix of the normal amplitude effects is diagonal with variances of the shifting and the scaling effects equal to $20^2$ and $0.05^2$ respectively.
  For the error process, the $\epsilon_{ij}$'s are normal white noise with variance, $\sigma^2_\epsilon = 5^2$ .
  Observations for each curve are sampled at equally spaced time points.  
  Three sampling frequencies $(n)$ are considered: 100, 1000 and 2000 points per curve.
  We simulate 20 curves for each simulation run with 200 runs for each setting.
  Figures~\ref{fig:shape-1} and~\ref{fig:shape-2} show the simulated curves of shape 1 with 100 points per curve and shape 2 with 1000 points per curve respectively.

  We compare the flexible SIM approach of \citet{brumback2004self}, which we refer to as BL2004, the phase and amplitude varying population pattern (PAVPOP) model of \citet{raket2014nonlinear}, the Bayesian hierarchical curve registration (BHCR) of \citet{telesca2008bayesian} and our proposed model and SAEM algorithm.
  We implemented our method in R and C++.
  R packages and source codes for BL2004 and BHCR are obtained from the authors whereas the R package for PAVPOP is available at the gitHub site: https://github.com/larslau/pavpop.
  While the four methods model the random effects differently, in our simulations, the basis functions to model the base shape and the warping functions are correctly specified for all methods.

  The R code for BL2004 and PAVPOP has built-in monitoring of convergence, however, no such functionality is implemented for SAEM and BHCR.
  Determining convergence at successive iterations appears to be an open question for SAEM and BHCR. 
  For our simulation study, we run a fixed schedule of 10000 iterations after 2000 burn-in iterations for these two methods.
  During the burn-in stage for SAEM, we fix the step-size of the stochastic approximation at $\gamma_k = 1$ to encourage the algorithm to move quickly to the region of high likelihood.
  We also tune the scale parameter of the Metropolis-Hastings proposal so that the acceptance rate is between 17\% and 33\%.

  The simulations are run on the Cedar cluster of Compute Canada.
  Table~\ref{tab:computing_time} describes the computing times for one successful simulation run and shows the percentage of runs that exceeded the maximum time allocated or failed due to numerical degeneracy.
  In terms of stability, the SAEM and BHCR approaches always return a fitted model whereas BL2004 and PAVPOP, which both rely on the approximation method of Lindstrom and Bates (1990) for a nonlinear mixed-effect model, can run into numerical errors.
  In most cases, the numerical errors are caused by degeneracy of matrices that are required to be positive definite. 
  For these two methods, the numerical problem is more prevalent for shape 2, possibly because the base shape is more complicated and the splines used in fitting the model are more flexible. 
  As the sampling frequency, $n$, increases, the failure rate for PAVPOP decreases while that for BL2004 increases.

  Comparing the computing time of successful runs when a fitted model is returned, SAEM is the fastest, while BHCR takes 2 to 3 times longer.
  This is likely due to the difference in the MCMC sampler of the two approaches.
  While BHCR updates the warping coefficients one at a time, our approach updates the entire vector of warping coefficients in a Metropolis-Hastings step.
  The simultaneous updating minimizes the number of times we need to reevaluate the basis functions of the common shape at the new warped time.
  Our method is therefore more scalable with the sampling frequency and the number of knots for the warping functions.
  On the other hand, both methods have low variability in their computing time since they are set up to run the same number of iterations.
  We would not consider BL2004 in this comparison as there was only one successful run, for shape 1 with $n=2000$.
  Among the remaining three methods, PAVPOP is the slowest because of the need to invert correlation matrices of the same dimension as the sampling frequency due to the Gaussian process assumption on the error process.
  The computing time is prohibitively long at high sampling frequency.
  In the case of shape 1 with $n=2000$, $92\%$ of the runs exceeded the maximum allocated time of 6 hours.

\begin{figure}[b]
    \centering
    \includegraphics[width=0.9\textwidth]{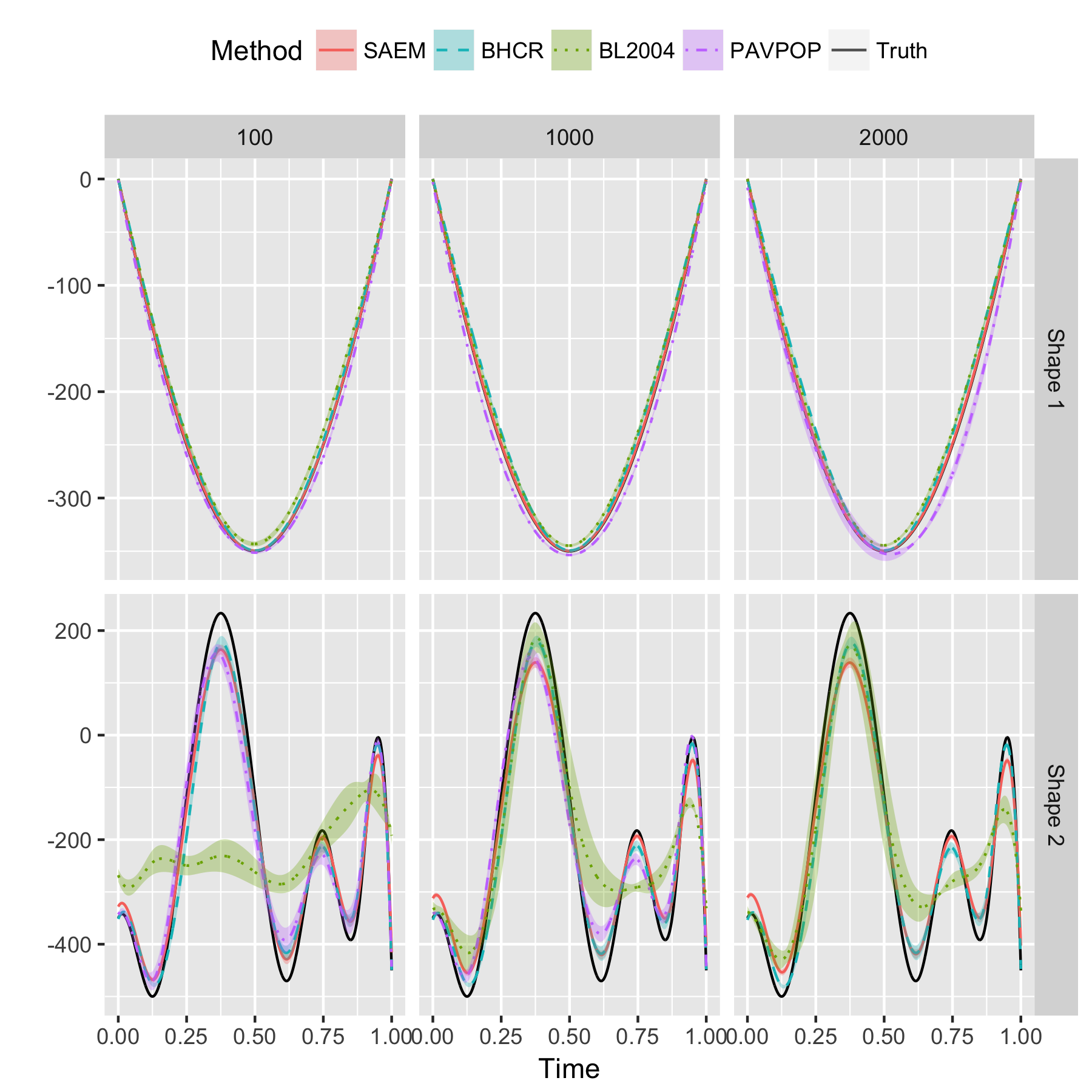}
    \caption{Average estimated common shape and pointwise 95\% confidence band based on the empirical standard error.}
    \label{fig:shape}
  \end{figure}

\begin{figure}[b]
    \centering
    \includegraphics[width=0.9\textwidth]{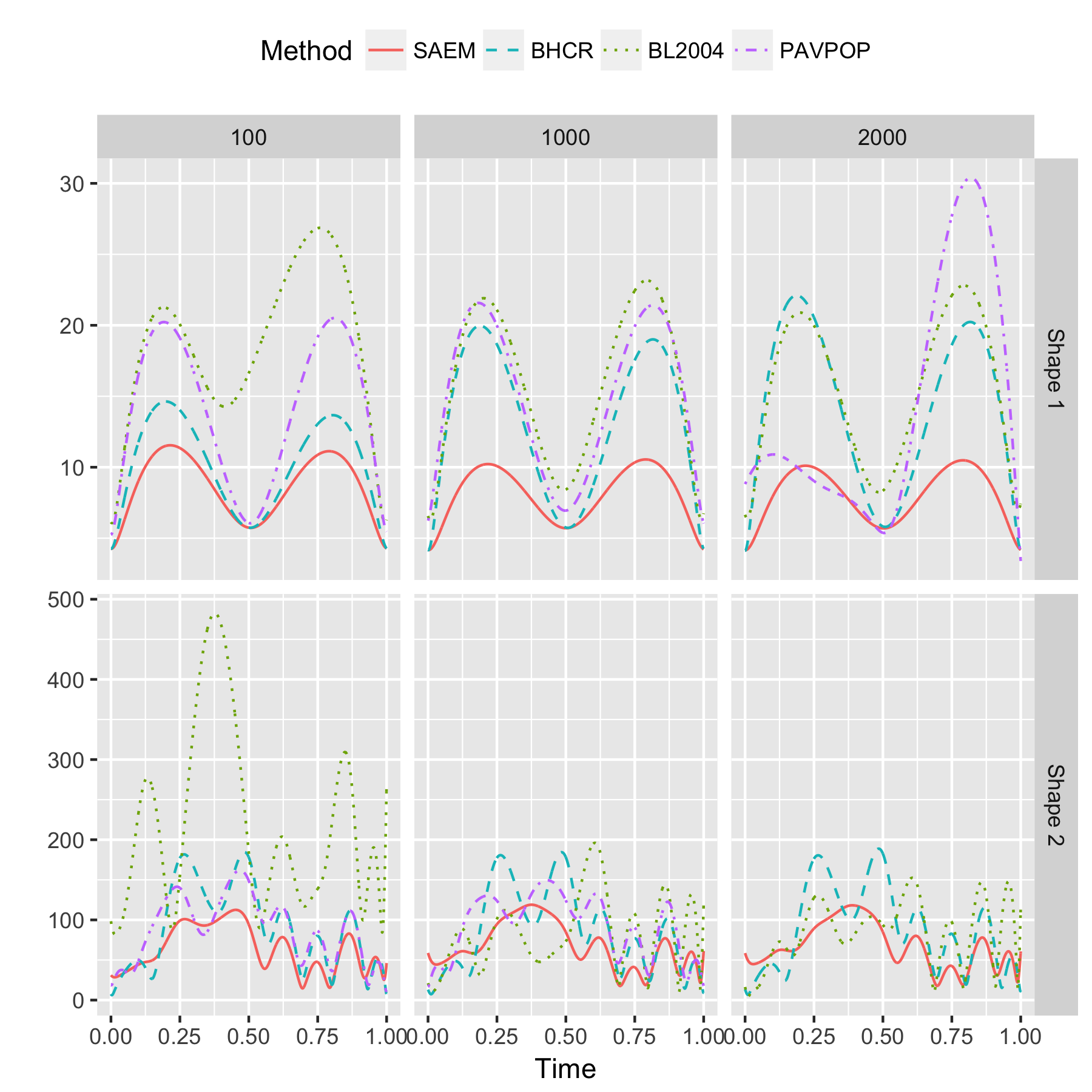}
    \caption{Root mean squared error of the estimated common shape.}
    \label{fig:RMSE}
  \end{figure}

  For estimating the base shape, all methods recover both shape 1 and shape 2 quite faithfully on average, as seen in Figure~\ref{fig:shape}.
  An exception is BL2004 which failed to reveal the smaller peaks and valleys of shape 2.
  Figure~\ref{fig:RMSE} shows the root mean squared errors (RMSEs) while Table~\ref{tab:average_imse} lists the integrated mean squared errors (IMSEs). 
  Our proposed method has the smallest IMSE in all cases while BHCR has the second lowest IMSE for shape 1.
  For BL2004, although convergence is not declared in most cases, the IMSEs are comparable to other methods, especially for shape 2, except in the case with $n=100$ where the bias is also substantial.

  We also compared the predicted warping functions.
  For SAEM and BHCR, the prediction bias and variability is modest in all cases.  
  The bias for PAVPOP is also small when it returns a fitted model.
  For BL2004, the averge prediction bias is much higher than the other methods for both shapes.
  In general, our method has the smallest integrated mean squared prediction error (IMSPE) of the predicted warping functions in the case of shape 1 while BHCR has the smallest IMSPE in the case of shape 2, as shown in Table~\ref{tab:average_imse}.

\section{Applications} 
\label{sec:applications}

  We apply our proposed method along with several other methods to a clustering analysis of marine mammal dive profiles (depth as a function of time during a single dive).  In one method, we cluster directly on the dive profiles using a technique of \citet{brillinger1997elephant}.  In another method, we  analyze the dive profiles with our curve registration method and cluster via the resulting warping functions.  As a third method, we  cluster using warping functions from the curve registration of  \citet{telesca2008bayesian}.  Clustering using warping functions appears to be superior to clustering on the dive profiles, and using warping functions from our method appears to be superior to using warping functions of \citet{telesca2008bayesian}.

  Dive depth profiles are one type of data that biologists use to study the diving activity of marine mammals, in order to understand the foraging areas and behaviour of the species~\citep{dragon2012horizontal,bailleul2008assessment,hindell1991diving}.  Studying diving activity can also help biologists assess the impact of environmental changes and human activities on the animals~\citep{guinet2014southern,walker2011classification} to inform the development of conservation policy.

\begin{figure}[b]
    \begin{subfigure}[b]{0.79\textwidth}
      \centering
      \includegraphics[width=0.95\textwidth]{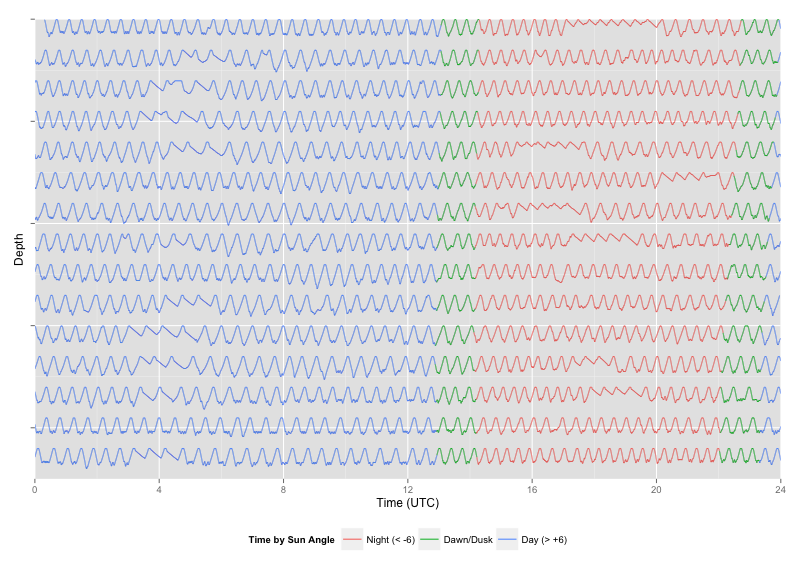}
      \caption{Daily time series of depth}
      \label{fig:daily-series}
    \end{subfigure}
    \begin{subfigure}[b]{0.49\textwidth}
      \centering
      \includegraphics[width=0.95\textwidth]{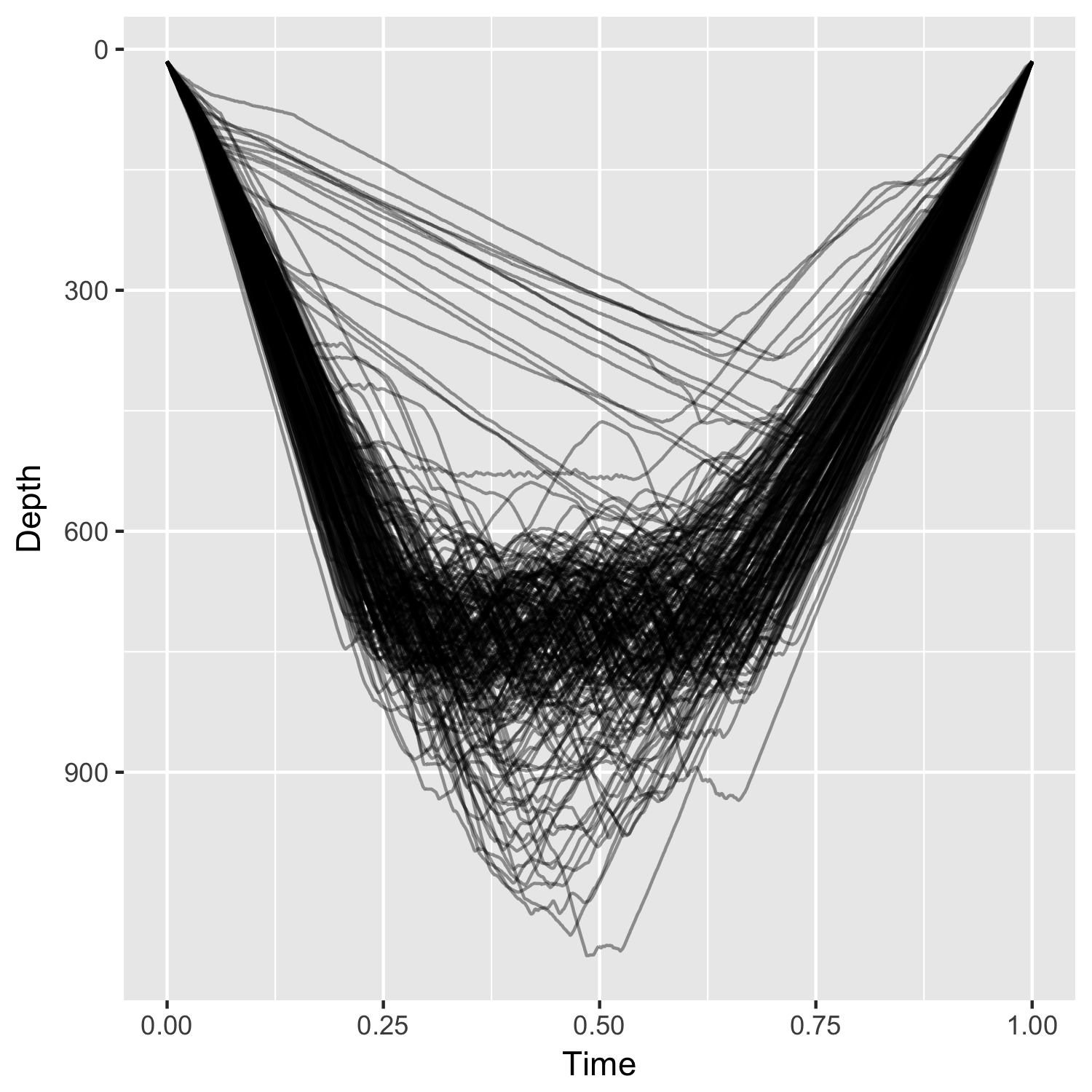}
      \caption{A random sample of 200 dive profiles}
      \label{fig:dive-profiles}
    \end{subfigure}
    \begin{subfigure}[b]{\textwidth}
      \centering
      \includegraphics[width=\textwidth]{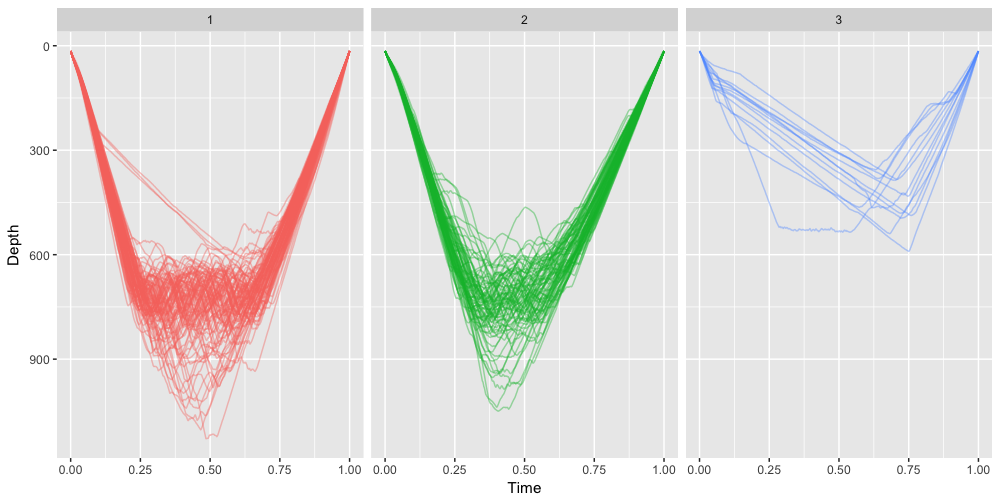}
      \caption{Unaligned dive profiles from clustering using the mixture model proposed by \cite{brillinger1997elephant}}
      \label{fig:brillinger_cluster}
    \end{subfigure}
    \caption{Dive profiles of a southern elephant seal}
    \label{fig:data}
  \end{figure}

  Since direct observation of underwater activity is challenging, researchers rely on various miniature sensors attached to the animals to collect proxy data such as location, movement, stomach temperature and ambient environmental parameters. 
  One basic device commonly deployed is the time depth recorder which tracks the dive depth of the tagged animal.
  Figure~\ref{fig:daily-series} shows the dive depth of a female southern elephant seal tagged on the Kerguelen Islands, recorded over a 61 day foraging trip.
  Depth is recorded at a per second frequency, yielding a high volume of data.
  From the figure, we can see that the seal dived repeatedly; each dive is about 20 to 30 minutes with short recesses of about 3 minutes at the surface between dives.

  A commonly adopted behavioural unit is dive.
  For each dive, the two dimensional time-depth trajectory is referred to as the dive profile.
  In this study, we analyze a sample of 200 randomly chosen dive profiles from the seal.
  Figure~\ref{fig:dive-profiles} shows the data; time is scaled such that dive durations are standardized to 1.
  We can see a few typical shapes among these curves.
  Physiological functions and behavioural states are often inferred based on the shape of the dive profile.

  While various multivariate clustering methods are proposed which cluster the dives based on summaries of the dive profiles, \citet{brillinger1997elephant} are perhaps the first to propose a model for the entire dive profile.
  They model dive profiles as a mixture of $M$ curves with dive profiles $f_1,\ldots, f_M$
  \begin{equation}
    P(m_i = k) = \pi_k, ~~ y_{ij} = f_{m_i}(t_{ij}) + \epsilon_{ij}
    \label{eq:brillinger-mixture}
  \end{equation}
  for $k = 1, \ldots, M$, and describe an easy to implement EM algorithm for estimation of the $f_k$'s and prediction of the cluster membership $m_i$.
  As our first clustering analysis, we use the method of \citet{brillinger1997elephant} to fit a 3-component mixture model to the 200 unaligned dive profiles.
  Figure~\ref{fig:brillinger_cluster} groups the dive profiles by the clusters identified.
  While the third cluster captures most of the drift dives with slower descent rates and shallower depths, the rest of the dives do not seem to be well separated into groups with distinctive shapes.
  Also, some dives also look out of place in their assigned cluster.

  Our preferred clustering analysis does not use the dive profile directly, but rather the predicted warping functions associated with the dive profiles. 
  We note that, in general, all dive profiles have common features: a descent stage, an ascent stage and a bottom stage with varying proportions of time in each stage across dives.
  Looking past the wiggles in the bottom stage, the different dive shapes can be viewed as warpings of a common bowl-shape curve up to a scaling factor in depth.
  Within the same dive type, warping functions will be similar to account for the similar proportions of times in descent, bottom and ascent phases of the dives.
  Between dive types, warping functions will differ more substantially to account for the systematic differences in their shapes.
  For this reason, we anticipate that a clustering analysis of the dive profiles based on phase variation would yield better results.
  We will align the 200 dive profiles and extract the amplitude and phase variations and then perform a clustering analysis on the predicted warping function to group the corresponding dive profiles into clusters of similar shape.  
  We use two methods of alignment, our SAEM method and the BHCR method of \citet{telesca2008bayesian}.

\begin{figure}[b]
    \centering
    \begin{subfigure}[b]{0.49\textwidth}
      \centering
      \includegraphics[width=0.95\textwidth]{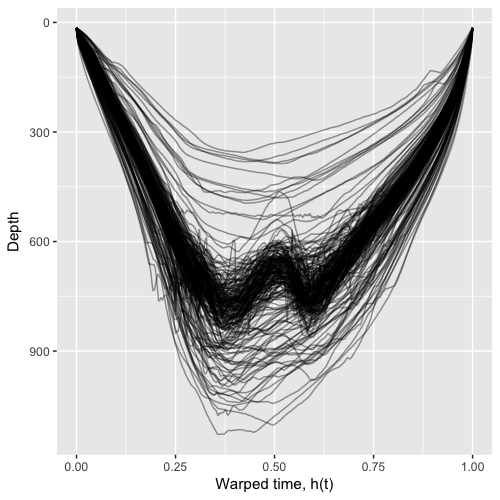}
      \caption{Aligned dive profiles}
      \label{fig:saem-aligned-profiles}
    \end{subfigure}
    \begin{subfigure}[b]{0.49\textwidth}
      \centering
      \includegraphics[width=0.95\textwidth]{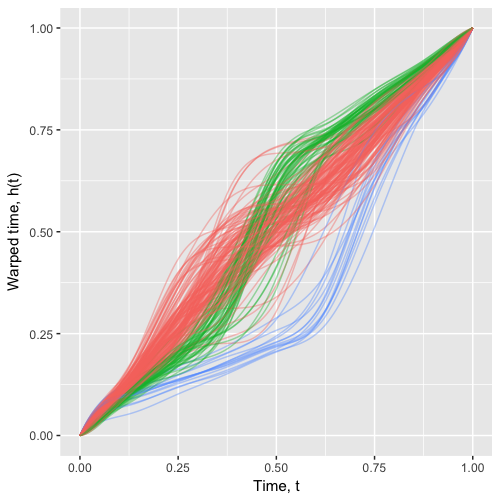}
      \caption{Estimated warping functions}
      \label{fig:saem-warping-functions}
    \end{subfigure}
    \begin{subfigure}[b]{\textwidth}
      \centering
      \includegraphics[width=\textwidth]{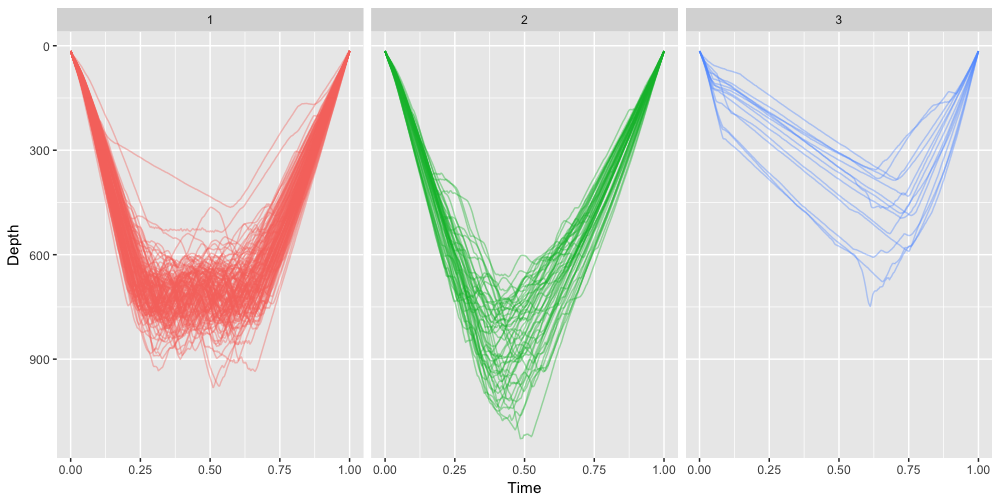}
      \caption{Unaligned dive profiles from clustering by K-means algorithm on basis function coefficients of the estimated warping functions.}
      \label{fig:saem-kmeanWarping}
    \end{subfigure}
    \caption{SAEM}
    \label{fig:saem-dive-profile}
  \end{figure}

  Figure~\ref{fig:saem-aligned-profiles} shows the aligned dive profiles from our SAEM method and Figure~\ref{fig:saem-warping-functions} shows the predicted warping functions.
  Since the predicted warping functions appear to follow three typical shapes, we group them into three clusters by a K-means algorithm on their basis coefficients.
  Figure~\ref{fig:saem-kmeanWarping} shows the dive profiles corresponding to the three clusters.
  We find that by extracting and clustering on the phase variations of the dive profiles, we achieve a better separation of the dives into W-shape dives with a longer sojourn at some target depth in the bottom stage, V-shape dives with negligible bottom stage, and drift dives with a much slower descending segment.

\begin{figure}[b]
    \centering
    \begin{subfigure}[b]{0.49\textwidth}
      \centering
      \includegraphics[width=0.95\textwidth]{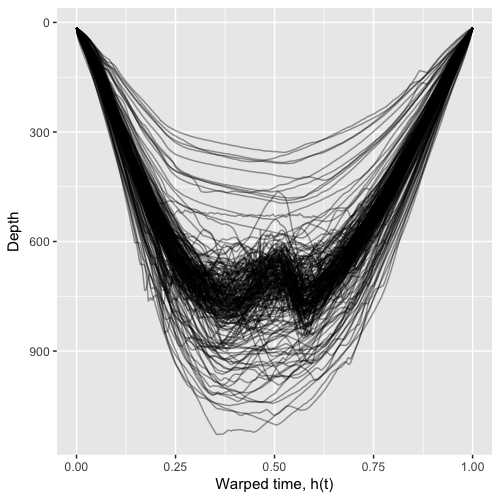}
      \caption{Aligned dive profiles}
      \label{fig:bhcr-aligned-profiles}
    \end{subfigure}
    \begin{subfigure}[b]{0.49\textwidth}
      \centering
      \includegraphics[width=0.95\textwidth]{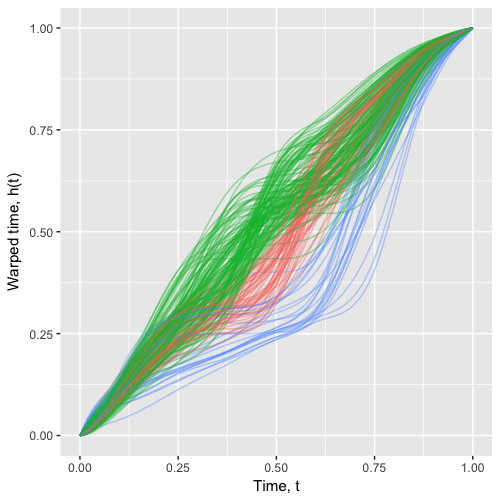}
      \caption{Estimated warping functions by BHCR}
      \label{fig:bhcr-warping-functions}
    \end{subfigure}
    \begin{subfigure}[b]{\textwidth}
      \centering
      \includegraphics[width=\textwidth]{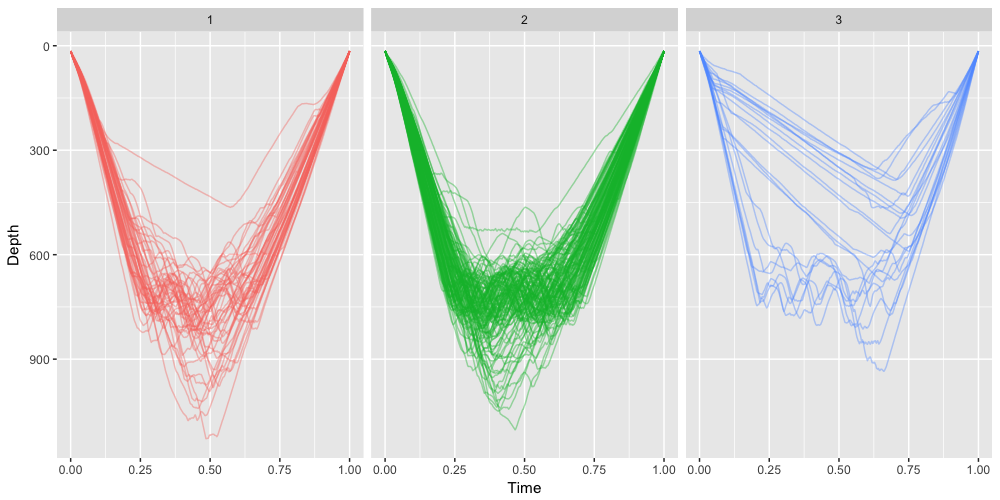}
      \caption{Unaligned dive profiles from clustering by K-means algorithm on basis function coefficients of the warping functions.}
      \label{fig:bhcr-kmeanWarping}
    \end{subfigure}
    \caption{BHCR}
    \label{fig:bhcr-dive-profile}
  \end{figure}

  The results from  the same K-means clustering analysis of the predicted warping functions from the BHCR analysis are shown in Figures~\ref{fig:bhcr-dive-profile} to \ref{fig:bhcr-kmeanWarping}.
  The dive profiles, however, are not as cleanly separated upon visual inspection.

\section{Discussion} 
\label{sec:discussion}

  In this article, we have discussed four model-based methods to analyze functional data with both amplitude variation and phase variation, namely the shape-invariant model with flexible time transformation of \citet{brumback2004self} (BL2004), the phase and amplitude varying population pattern (PAVPOP) model of \citet{raket2014nonlinear}, the Bayesian hierarchical curve registration (BHCR) model of \citet{telesca2008bayesian} and our new method which estimates a curve registration model similar to BL2004 by a stochastic approximation EM (SAEM) algorithm.

  Based on simulation studies, we find that BL2004 and PAVPOP often run into numerical issues.
  When the number of points sampled per curve is high, PAVPOP is computationally inefficient, whereas BL2004 is particularly unstable and can produce poor fit when the curves are multimodal.

  On the other hand, SAEM and BHCR are more stable and always produce an estimate.
  We obtained good fits to the data in comparable computing time using the two methods, with SAEM faster in general and more scalable in time with respect to points per curves and flexibility of the warping function.
  For estimating the base shapes and predicting the warping functions, both SAEM and BHCR have negligible bias and similar IMSE.

  We then applied SAEM and BHCR in a clustering analysis of southern elephant seal dive profiles.
  While both methods appear to align the dive profiles well, when we cluster the profiles by applying the K-means algorithm to their associated warping functions, the warping functions predicted by SAEM generated visually more meaningful clusters.

\section*{Acknowledgements} 
\label{sec:acknowledgement}
  We would like to thank Dr. Christophe Guinet of Centre d'Etudes Biologiques de Chiz\'{e} UMR 7372 CNRS-ULR for providing us with the Southern Elephant Seal data. The data were collected as part of the Syst\`{e}me d'Observation MEMO.
  We also thank Victor Panaretos and Ed Perkins for helpful discussions.
  This work was supported by the National Science and Engineering Research Council of Canada, grant number 7969.  

\appendix

\section{Proof of Theorem \ref{thm:identifiable}}

  The proof of the theorem requires some  topological arguments because of requirements of measurability of mappings between function spaces and measurability of subsets of function spaces.  We provide details here.
  We make full use of the following argument.  Suppose $\Omega_i$, $i=1, 2$,  are metric spaces with corresponding Borel sigma-algebras ${\cal{O}}_i$ and  $\cal{H}$ is a  function from $\Omega_1$ to $\Omega_2$.   Let $(\Omega,{\mathcal{F}}, P)$ be a probability space. 
  If  $X_1$ and $X_2$  are random variables from $\Omega$ to $\Omega_1$ with  
  $P\{ X_1 \in B \} = P\{ X_2 \in B \}$ for all $B \in {\cal{O}}_1$, then   
  $P\{ {\mathcal{H} }(X_1) \in C \} = P\{{\mathcal{H} } (X_2) \in C \}$ for all $C \in {\cal{O}}_2$
  provided  that ${\mathcal{H}}^{-1}(C) \in {\cal{O}}_1$ for all $C \in {\cal{O}}_2$, that is, provided that ${\mathcal{H}}$ is measurable. 
  In our proofs, we  show that the function
  ${\cal{H}}$ is measurable by showing that    ${\cal{H}}$ is
  continuous with respect to the metrics on $\Omega_1$ and $\Omega_2$.
  In some cases, the function ${\mathcal{H}}$ will only be defined on ${\mathbb{A}}$, a subset of $\Omega_1$.  In this case, we show that ${\mathbb{A}}$ is in ${\mathcal{O}}_1$ and ${\mathcal{H}}$ is continuous on ${\mathbb{A}}$.

  We define the following function spaces and distances, which induce topologies.
  \begin{itemize}
    \item $ C^1[0,1]$ is the set of functions from $[0,1] \to \Re$ with continuous first derivatives, with distance 
    \[
    d_{\inftystar}(f_1,f_2) \equiv || f_1-f_2||_\inftystar  \equiv \sup_t |f_1(t) - f_2(t)|  + \sup_t |f_1'(t) - f_2'(t)|.
    \]
    \item $C[0,1]$ is the set of  continuous functions from $[0,1] \to \Re$,  with distance
    \[
    d_\infty(f_1,f_2) \equiv  || f_1-f_2||_\infty = \sup_t | f_1(t)-f_2(t)| .
    \]
  \end{itemize}
  For any cross-products of function spaces, we use the usual cross-product metric/topology.  
  The following lemmas are presented without proof.
 
  \begin{lemma}
  \label{lemma:sets}
  \hfill\break
   \begin{enumerate}
   \item
   Let  ${\mathbb{A}} \subseteq C^1[0,1]$ consist of all non-constant functions.    In the topology induced by $d_\inftystar$, $\mathbb{A}$ is an open set.
   \vskip 10pt
          \item 
        Let ${\mathcal{B}}\subseteq C^1[0,1]$ with ${\mathcal{B}} $ consisting of $h \in C^1[0,1]$ with $h(0)=0$, $h(1)=1$ and $h'(t)>0$ for all $t$.  The set ${\mathcal{B}}$ is a Borel set of $C^1[0,1]$ in the topology induced by $d_\inftystar$.
        \vskip 10pt
       \item
       Let ${\mathcal{C}} \subseteq C^1[0,1]$ $=\{ F \in C^1[0,1]: F(0)=0, F(1)=1, F(t) > F(s)$ for all $t>s\}$.  Then ${\mathcal{C}}$ is a measurable subset of $C^1[0,1]$ in the topology induced by $d_\inftystar$.
     \end{enumerate}  
  \end{lemma}
  \vskip 10pt

  \begin{lemma}
  \label{lemma:maps}
  The following mappings are well-defined and continuous, with the topologies on ${\mathbb{A}}, {\mathcal{B}}, {\mathcal{C}}$ and $C^1[0,1]$ induced by $d_\inftystar$ and the topology on $C[0,1]$ induced by the usual $L_\infty$ norm.
  \hfill\break
  \begin{enumerate}
  \item    $h \in {\mathcal{B}} \to h^{-1} \in {\mathcal{B}}$;
  \vskip 10pt
  \item
   $ {\mathcal{H}}_{\circ}:
  C^1[0,1] \times C^1[0,1]  \to C^1[0,1] $ with
  ${\mathcal{H}}_{\circ}(f,g)  = f\circ g; $
  \vskip 10pt
  \item $ {\mathcal{H}}_1  : C^1[0,1] \times C^1[0,1]   \to C^1[0,1] \times C^1[0,1]$ with
      ${\mathcal{H}}_1 (f,h) = (f \circ h, h)$;
  \vskip 10pt
  \item
  ${\mathcal{H}}_{\rm{inv}}:  {\mathcal{C}}  \to     C[0,1]$, with
  ${\mathcal{H}}_{\rm{inv}}( F)  =  F^{-1}$;
  \vskip 10pt
  \item  the total variation mapping ${\mathcal{H}}_{TV}: {\mathbb{A}} \to C^1[0,1]$:
  \begin{equation}
    ( { \mathcal{H}}_{TV} f) (t) = \frac{\int_{0}^{t}| f'(s)|\text{d}s}{\int_{0}^{1}|f'(u)|\text{d}u}.
    \nonumber
  \end{equation}
  \end{enumerate}

  \end{lemma}

  \begin{lemma}
  \label{lemma:equalities}
  The following equalities can be found in Lemma 1 of  \citet{chakraborty2017functional}.  
  \[
  F_i \equiv { \mathcal{H}}_{TV} (f_i)  = { \mathcal{H}}_{TV} (\xi_i) \equiv F_{\xi_i},
  \] 
  \begin{equation}
  \tilde{F}_i \equiv { \mathcal{H}}_{TV} (\tilde{f}_i)  ={ \mathcal{H}}_{TV} (\tilde{\xi_i}) \equiv \tilde{F}_{\xi_i}
  \nonumber
  \end{equation}
  and
   \begin{equation}
   \label{eq:F.xi}
  \tilde{F}_i  =  F_{\xi_i} \circ h_i.
   \end{equation}
  \end{lemma}

    \vskip20pt
    \noindent
    {\bf{Proof of the Theorem}}

       We use the notation introduced in Lemmas \ref{lemma:sets}, \ref{lemma:maps} and \ref{lemma:equalities}.
      Suppose that $(f_1,h_1) \distributedas (f_2,h_2)$.  Then  continuity of ${\mathcal{H}}_o$ implies that $f_1 \circ h_1 \distributedas f_2 \circ h_2$.

   Now we suppose that   $\tilde{f}_1 \distributedas \tilde{f}_2$ and show that    $ (f_1, h_1)  \distributedas (f_2, h_2)$.
   We proceed by showing 
    \begin{enumerate}[(i)]
   \item   $(\tilde{f}_1,\tilde{F}_1) \distributedas (\tilde{f}_2,\tilde{F}_2)$, 
  \item   $\tilde{F}_1^{-1} \distributedas \tilde{F}_2^{-1}$,   
  \item  $ {F}_{\xi_1} = {F}_{\xi_2}$,
  and finally
  \item  $(f_1,h_1) \distributedas (f_2,h_2)$.
  \end{enumerate}

   We see that  (i) follows by the continuity of  the mapping ${ \mathcal{H}}_{TV} $  and the measurability of the set ${\mathbb{A}}$.
   Item  (ii) follows since $\tilde{F}_i \in {\mathcal{C}}$ and the mapping ${\mathcal{H}}_{\rm{inv}}$ is continuous on ${\mathcal{C}}$, which is a measurable set.

  \vskip 15pt
   
   To show  (iii),  we note that $\tilde{F}_i^{-1}(t)$ is bounded and so its expectation exists.  By (ii),
   \[
  \E[ \tilde{F}_1^{-1}(t)]  = \E[ \tilde{F}_2^{-1}(t)] ~~{\rm{~for~all~}}t.
  \]
   From equation (\ref{eq:F.xi}), $\tilde{F}_i^{-1}  =  h_i^{-1} \circ F^{-1}_{\xi_i}$, so
   \[
  \E[ \tilde{F}_i^{-1}(t)] =  \E[ h_i^{-1}({F}_{\xi_i}^{-1}(t))]    
  =
  {F}_{\xi_i}^{-1}(t) 
  \]
  by  the fact that ${F}_{\xi_i}^{-1}(t)$ is nonrandom  and the assumption that $\E[h_i^{-1}(u)]=u$ for all $u$. 
  Therefore, ${F}_{\xi_1}^{-1} = {F}_{\xi_2}^{-1}$ and (iii) holds.

   \vskip 15pt

   To show  (iv), we write, for $A$ in the sigma-algebra associated with ${\mathbb{A}}$ 
   and $B $ in the sigma-algebra associated with $ {\mathcal{B}}$, by continuity of the mapping  ${\mathcal{H}}_1$.
   \begin{eqnarray}
   P\{ (f_1,h_1)  \in A \times B \}  &=&
   P\{ (f_1 \circ h_1, h_1)  \in  {\mathcal{H}}_1( A \times B) \} 
   \nonumber \\
  &= &
  \label{eq:composition}
   P\{ (f_1 \circ h_1, {F}_{\xi_1} \circ h_1)  \in ( {\mathcal{H}}_2 \circ {\mathcal{H}}_1)( A \times B) \}
  \end{eqnarray}
  where
  $
   {\mathcal{H}}_2 :  {\mathbb{A}} \times {\mathcal{B}} \to  {\mathbb{A}}  \times C^1[0,1]
  $ mapping
  $ (f,h) $  to $ (f, F_{\xi_1} \circ h)$
  is continuous since the composition of functions is continuous.
  But, by (\ref{eq:F.xi}),  expression (\ref{eq:composition}) is equal to
  \[
    P\{ (\tilde{f}_1,  \tilde{F}_1)  \in  ( {\mathcal{H}}_2 \circ {\mathcal{H}}_1)( A \times B) \} 
   \]
   which, by (i), equals
   \[ 
    P\{ (\tilde{f}_2,  \tilde{F}_2)  \in  ( {\mathcal{H}}_2 \circ {\mathcal{H}}_1)( A \times B) \} .
    \]
  Similarly, 
  \begin{eqnarray}
  P\{ (f_2 , h_2) \in A \times B \} 
   & =&
     P\{ (f_2 \circ h_2, {F}_{\xi_1} \circ h_2)  \in ( {\mathcal{H}}_2 \circ {\mathcal{H}}_1)( A \times B) \}
     \nonumber \\
    & =&
       P\{ (f_2 \circ h_2, {F}_{\xi_2} \circ h_2)  \in ( {\mathcal{H}}_2 \circ {\mathcal{H}}_1)( A \times B) \}  {\rm{~~~~by~(iii)}}
     \nonumber \\
    &=&   P\{ (\tilde{f}_2, \tilde{F}_2)  \in ( {\mathcal{H}}_2 \circ {\mathcal{H}}_1)( A \times B) \}.
    \nonumber
     \end{eqnarray}
    Thus $(f_1,h_1)  \distributedas (f_2, h_2)$.

  \FloatBarrier

\begin{table}[ht]
  \centering
  \caption{Computing time for one successful simulation run and the percentage of failed runs for varying number of sampled points. We allocate a maximum computing time of 6 hours for each run.}
  \label{tab:computing_time}
  \begin{tabular}{lllrrrrrr}
  \hline
  & & & \multicolumn{2}{c}{Times (sec)} 
  & \multicolumn{2}{c}{Aborted Runs} 
  & \multicolumn{2}{c}{Compeleted Runs} \\
  \cline{4-5} \cline{6-7} \cline{8-9}
  Shape & n & Method & Median & IQR & 
  \multicolumn{1}{p{1.5cm}}{Max. time exceeded} & 
  \multicolumn{1}{p{1.5cm}}{Numerical error encountered} & 
  \multicolumn{1}{p{1.5cm}}{non-convergence declared} &
  \multicolumn{1}{p{1.5cm}}{No errors or warning}\\ 
  \hline
    \multirow{12}{*}{1} & \multirow{4}{*}{100}&
       SAEM       &    19.5   &     0.9   &   0.0\%   &     0.0\%  &      -  &   100\%\\
    && BHCR       &    46.0   &     4.8   &   0.0\%   &     0.0\%  &      -  &   100\%\\
    && BL2004     &    65.8   &    81.1   &   0.0\%   &    21.5\%  & 78.5\%  &   0.0\%\\
    && PAVPOP     &   169.6   &    53.1   &   0.0\%   &    40.5\%  &  3.5\%  &  56.0\%\\
    \cline{2-9}
    & \multirow{4}{*}{1000}&
       SAEM       &   151.0   &     1.2   &   0.0\%   &     0.0\%  &      -  &   100\%\\
    && BHCR       &   421.3   &     7.5   &   0.0\%   &     0.0\%  &      -  &   100\%\\
    && BL2004     &   235.4   &   255.8   &   0.0\%   &    41.0\%  & 59.0\%  &   0.0\%\\
    && PAVPOP     &  7541.1   &  2399.7   &   0.0\%   &    11.5\%  &  7.5\%  &  81.0\%\\
    \cline{2-9}
    & \multirow{4}{*}{2000}&
       SAEM       &   295.3   &    26.7   &   0.0\%   &     0.0\%  &      -  &   100\%\\
    && BHCR       &   881.6   &    15.6   &   0.0\%   &     0.0\%  &      -  &   100\%\\
    && BL2004     &   271.3   &   360.2   &   0.0\%   &    46.0\%  & 53.5\%  &   0.5\%\\
    && PAVPOP     & 19480.6   &  2517.8   &  92.0\%   &     6.5\%  &  0.0\%  &   1.5\%\\
    \hline
    \multirow{12}{*}{2} & \multirow{4}{*}{100}&
       SAEM       &    30.0   &     0.8   &   0.0\%   &     0.0\%  &      -  &   100\%\\
    && BHCR       &    71.4   &     2.6   &   0.0\%   &     0.0\%  &      -  &   100\%\\
    && BL2004     &   386.0   &    24.3   &   0.0\%   &    63.5\%  & 36.5\%  &   0.0\%\\
    && PAVPOP     &   612.9   &   247.1   &   0.0\%   &    72.5\%  &  3.5\%  &  24.0\%\\
    \cline{2-9}
    & \multirow{4}{*}{1000}&
       SAEM       &   242.8   &     3.9   &   0.0\%   &     0.0\%  &      -  &   100\%\\
    && BHCR       &   711.9   &    28.7   &   0.0\%   &     0.0\%  &      -  &   100\%\\
    && BL2004     &   288.6   &   137.0   &   0.0\%   &    98.0\%  &  2.0\%  &   0.0\%\\
    && PAVPOP     & 12668.4   &  3248.6   &   0.0\%   &    14.5\%  &  3.5\%  &  82.0\%\\
    \cline{2-9}
    & \multirow{4}{*}{2000}&
       SAEM       &   499.2   &    14.7   &   0.0\%   &     0.0\%  &      -  &   100\%\\
    && BHCR       &  1756.4   &    79.5   &   0.0\%   &     0.0\%  &      -  &   100\%\\
    && BL2004     &   490.2   &   207.8   &   0.0\%   &    97.5\%   & 2.5\%  &   0.0\%\\
    && PAVPOP     &       -   &       -   &  73.5\%   &    26.5\%   & 0.0\%  &   0.0\%\\
    \hline\hline
  \end{tabular}
  \end{table}
  \normalsize

  \begin{table}[b]
  \centering
  \caption{Average IMSE of the estimated common shape and IMSPE of the predicted warping function}
  \label{tab:average_imse}
  \begin{tabular}{lllrr}
    \hline
    Shape & n & Method & \multicolumn{1}{p{1.2in}}{Average IMSE of common shape} & \multicolumn{1}{p{1.2in}}{Average IMSPE of warping functions}\\ 
    \hline
       \multirow{12}{*}{1} 
     & \multirow{4}{*}{100}
     & SAEM     &    79   & $  0.14 \times 10^{-3}$\\
    && BHCR     &   114   & $  0.20 \times 10^{-3}$\\
    && BL2004   &   379   & $  2.75 \times 10^{-3}$\\
    && PAVPOP   &   222   & $  0.52 \times 10^{-3}$\\
    \cline{2-5}
     & \multirow{4}{*}{1000}
     & SAEM     &    68   & $  0.11 \times 10^{-3}$\\
    && BHCR     &   198   & $  0.35 \times 10^{-3}$\\
    && BL2004   &   281   & $  2.22 \times 10^{-3}$\\
    && PAVPOP   &   250   & $  0.49 \times 10^{-3}$\\
    \cline{2-5}
     & \multirow{4}{*}{2000}
     & SAEM     &    68   & $  0.11 \times 10^{-3}$\\
    && BHCR     &   231   & $  0.39 \times 10^{-3}$\\
    && BL2004   &   266   & $  2.16 \times 10^{-3}$\\
    && PAVPOP   &   268   & $  0.52 \times 10^{-3}$\\
     \hline
       \multirow{12}{*}{2} 
     & \multirow{4}{*}{100}
     & SAEM     &  4807   & $  3.85 \times 10^{-3}$\\
    && BHCR     & 10677   & $  3.74 \times 10^{-3}$\\
    && BL2004   & 60333   & $181.82 \times 10^{-3}$\\
    && PAVPOP   &  9415   & $  6.12 \times 10^{-3}$\\
    \cline{2-5}
     & \multirow{4}{*}{1000}
     & SAEM     &  5322   & $  5.35 \times 10^{-3}$\\
    && BHCR     & 10088   & $  3.83 \times 10^{-3}$\\
    && BL2004   &  8812   & $ 15.45 \times 10^{-3}$\\
    && PAVPOP   & 10062   & $  4.68 \times 10^{-3}$\\
    \cline{2-5}
     & \multirow{4}{*}{2000}
     & SAEM     &  5256   & $  5.39 \times 10^{-3}$\\
    && BHCR     & 10638   & $  3.42 \times 10^{-3}$\\
    && BL2004   &  8640   & $ 15.02 \times 10^{-3}$\\
    && PAVPOP   &   -     &                       -\\
    \hline\hline
  \end{tabular}
  \end{table}

  \begin{figure}[b]
    \centering
    \includegraphics[width=0.9\textwidth]{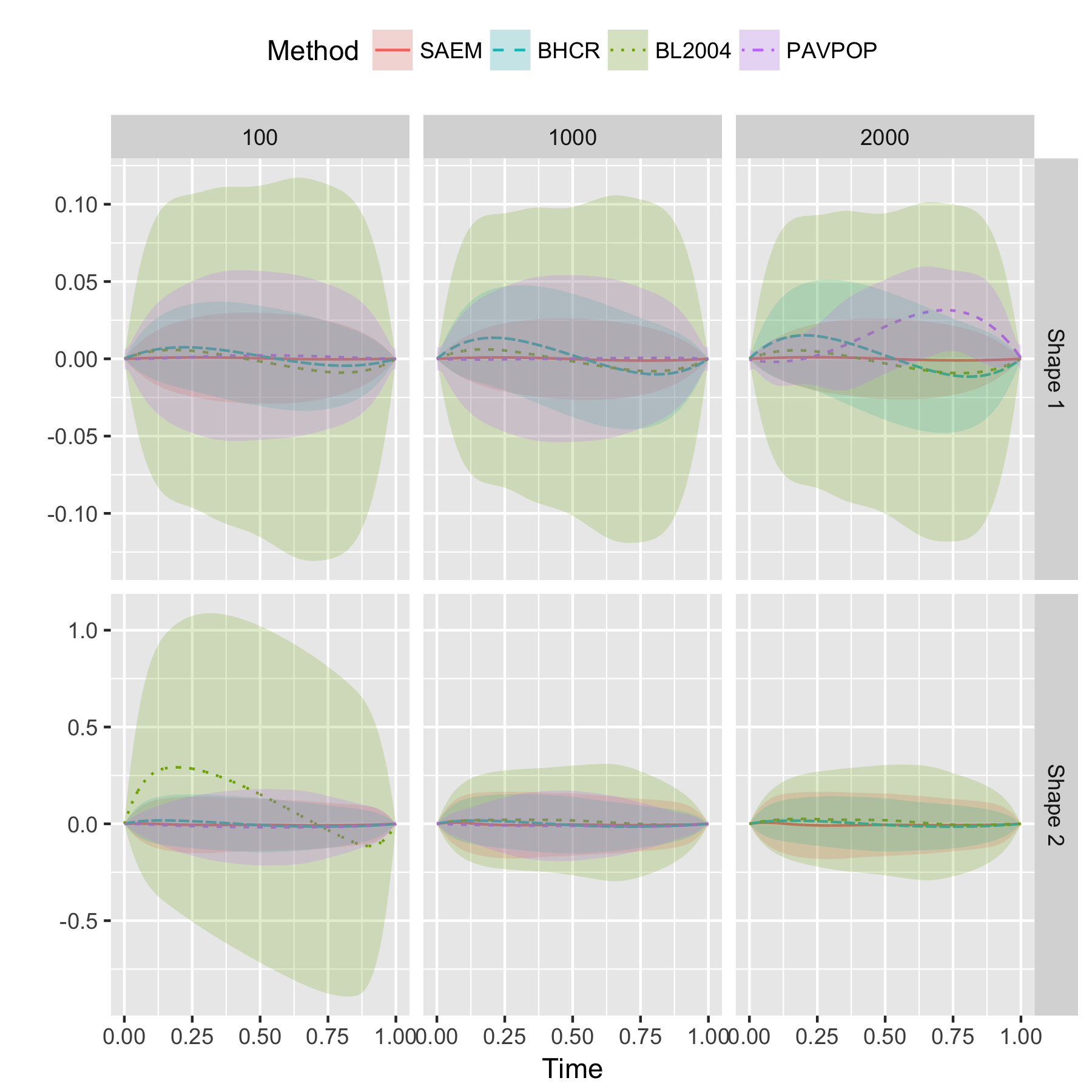}
    \caption{Average prediction error of the warping functions and pointwise 95\% confidence band based on the empirical standard error.}
    \label{fig:bias-warping}
  \end{figure}

  \begin{figure}[b]
    \centering
    \includegraphics[width=\textwidth]{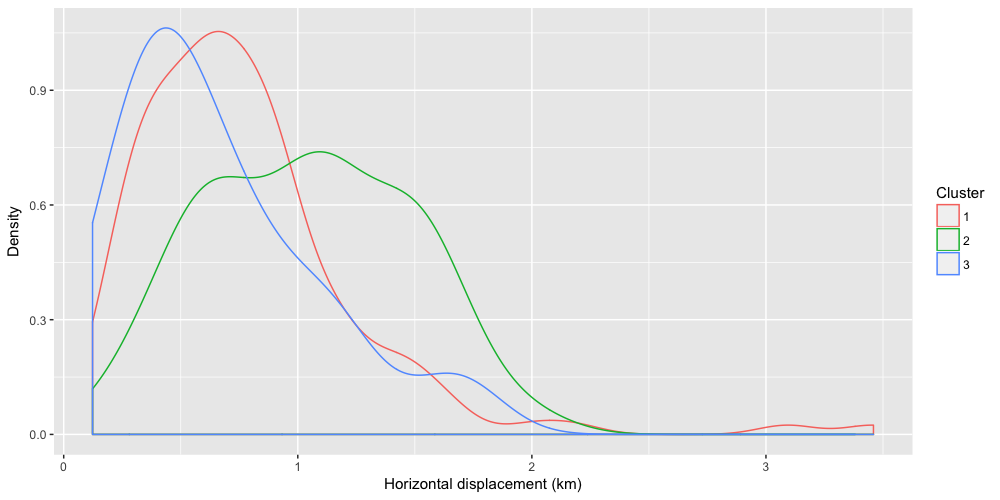}
    \caption{Kernel density estimates of horizontal displacement for each dive cluster identified by K-means algorithm on basis function coefficients of the estimated warping functions by our proposed method.}
    \label{fig:h_dist}
  \end{figure}

\FloatBarrier
\bibliography{reference}
\bibliographystyle{imsart-nameyear}

\end{document}